\documentclass{emulateapj}

\usepackage{xspace}
\usepackage{graphics,graphicx}
\usepackage{natbib}
\usepackage{multirow}
\usepackage[percent]{overpic}
\usepackage{color}

\shorttitle{Identifying planetary biosignature impostors}
\shortauthors{Schwieterman et al.}

\begin{document}

\title{Identifying planetary biosignature impostors:  Spectral features of CO and O$_{4}$ resulting from abiotic O$_{2}$/O$_{3}$ production}
\author{Edward W. Schwieterman\altaffilmark{1,2,3}, Victoria S. Meadows\altaffilmark{1,2,3}, Shawn D. Domagal-Goldman\altaffilmark{2,4}, Drake Deming\altaffilmark{2,5}, Giada N. Arney\altaffilmark{1,2,3}, Rodrigo Luger\altaffilmark{1,2,3}, Chester E. Harman\altaffilmark{2,6,7,8}, Amit Misra\altaffilmark{1,2,3}, Rory Barnes\altaffilmark{1,2,3}
}
\altaffiltext{1}{Astronomy Department, University of Washington, Box 351580, Seattle, WA 98195 USA, eschwiet@uw.edu}
\altaffiltext{2}{NASA Astrobiology Institute's Virtual Planetary Laboratory, Seattle, WA 981195 USA}
\altaffiltext{3}{Astrobiology Program, University of Washington, Seattle, WA 98195 USA}
\altaffiltext{4}{NASA Goddard Space Flight Center, Greenbelt, MD 20771 USA}
\altaffiltext{5}{Department of Astronomy, University of Maryland, College Park, MD 20742 USA}
\altaffiltext{6}{Geosciences Department, Pennsylvania State University, University Park, PA 16802, USA}
\altaffiltext{7}{Pennsylvania State Astrobiology Research Center, 2217 Earth and Engineering Sciences Building, University Park, PA 16802, USA}
\altaffiltext{8}{Center for Exoplanets and Habitable Worlds, Pennsylvania State University, University Park, PA 16802, USA}

\begin{abstract}
O$_{2}$ and O$_{3}$ have been long considered the most robust individual biosignature gases in a planetary atmosphere, yet multiple mechanisms that may produce them in the absence of life have been described.  However, these abiotic planetary mechanisms modify the environment in potentially identifiable ways. Here we briefly discuss two of the most detectable spectral discriminants for abiotic O$_{2}$/O$_{3}$: CO and O$_{4}$. We produce the first explicit self-consistent simulations of these spectral discriminants as they may be seen by \textit{JWST}. If \textit{JWST}-NIRISS and/or NIRSpec observe CO (2.35, 4.6 $\mu$m) in conjunction with CO$_{2}$ (1.6, 2.0, 4.3 $\mu$m) in the transmission spectrum of a terrestrial planet it could indicate robust CO$_{2}$ photolysis and suggest that a future detection of O$_{2}$ or O$_{3}$ might not be biogenic. Strong O$_{4}$ bands seen in transmission at 1.06 and 1.27 $\mu$m could be diagnostic of a post-runaway O$_{2}$-dominated atmosphere from massive H-escape. We find that for these Òfalse positiveÓ scenarios, CO at 2.35 $\mu$m, CO$_{2}$ at 2.0 and 4.3 $\mu$m, and O$_{4}$ at 1.27 $\mu$m are all stronger features in transmission than O$_{2}$/O$_{3}$ and could be detected with SNRs $\gtrsim$ 3 for an Earth-size planet orbiting a nearby M dwarf star with as few as 10 transits, assuming photon-limited noise.  O$_{4}$ bands could also be sought in UV/VIS/NIR reflected light (at 0.345, 0.36, 0.38, 0.445, 0.475, 0.53, 0.57, 0.63, 1.06, and 1.27 $\mu$m) by a next generation direct-imaging telescope such as LUVOIR/HDST or HabEx and would indicate an oxygen atmosphere too massive to be biologically produced.  
\end{abstract}
\keywords{astrobiology---planets and satellites: atmospheres---planets and satellites: terrestrial planets---techniques: spectroscopic} 


\section{Introduction}
\setcounter{footnote}{0}
In recent years dozens of exoplanets have been identified that are likely or confirmed to be rocky in composition, analogous to the solar system's inner terrestrial planets (e.g., \citealt{Berta-Thompson+15}). Some of these known planets reside within the Òhabitable zoneÓ (HZ) of their host star \citep{Kopparapu+13}, but are too distant from Earth for spectroscopic follow-up to characterize their planetary environments.  The upcoming \textit{Transiting Exoplanet Survey Telescope} (TESS; \citealt{Ricker+14,Sullivan+15}) and ground-based surveys (e.g., \citealt{Charbonneau+09}) will identify more nearby terrestrial HZ planets that are amenable to further study by missions such as the \textit{James Webb Space Telescope} (JWST; \citealt{Deming+09}) or large ground-based telescopes (e.g., \citealt{Snellen+13}).   Many of these terrestrial exoplanets will have secondary atmospheres in which photochemistry, climate, history of atmospheric escape, volcanic outgassing, and perhaps biology will play significant roles in their atmospheric compositions (e.g., Segura et al. 2005, 2010). Understanding these star-planet interactions and geological processes will be important for characterizing a terrestrial planet's atmosphere, both to understand the planetary environment, and to increase the confidence with which we can identify planetary phenomena as being more likely to be produced by life (e.g., \citealt{DesMarais+02,Segura+05, Seager+12}) than by abiotic processes.

Atmospheric O$_{2}$ (or its photochemical byproduct O$_{3}$) is often considered a robust astronomical biosignature, because through the history of our planet life has been the dominant source of this gas. The exact mechanism(s) and timeline for the oxygenation of EarthÕs atmosphere is still under  debate, but there is broad agreement on the fundamental causes \citep{Lyons+14}. The O$_{2}$ in Earth's atmosphere is unstable over geological timescales and would be depleted by reactions with reduced volcanic gases and through oxidation of the surface \citep{Catling13}, thus requiring an active source to maintain appreciable levels. On Earth, that active source exists in the form of photosynthetic production of O$_{2}$ by life, followed by burial of organic carbon \citep{Catling13}, which separates the organic carbon material from atmospheric O$_{2}$. The photochemical production of very small amounts of O$_{2}$ occurs on Earth from the photolysis of O-bearing molecules, but would not build up to appreciable levels in the absence of biology due to the shape of the UV spectrum of the Sun and significant sinks for O$_{2}$ \citep{Domagal-Goldman+14,Harman+15}. The O$_{3}$ in Earth's atmosphere is a result of photochemical reactions involving O$_{2}$, and the presence of large quantities of O$_{3}$ has been suggested as being an indicator of abundant, photosynthetically-generated O$_{2}$ \citep{DesMarais+02, Segura+05}. Recently, it has been calculated that the strongest disequilibrium indicator in Earth's atmosphere-ocean system in terms of free energy is the N$_{2}$-O$_{2}$-dominated atmosphere coexisting with a liquid H$_{2}$O surface ocean \citep{Krissansen-Totton+15}. The common thread to all of these signatures is the presence of photosynthetically-produced O$_{2}$, and there have been many studies into the feasibility of future ground-based and space-based observations to detect O$_{2}$ spectral features in exoplanet atmospheres (e.g., \citealt{Snellen+13, Misra+14a, Dalcanton+15}).

Even though abiotic sources for abundant O$_{2}$/O$_{3}$ are not present for Earth, three major categories of abiotic O$_{2}$ mechanisms have been recently identified that may affect planets with different atmospheric histories or within the HZs of different types of host star. These include: 1) the photochemical production of stable concentrations of  O$_{2}$/O$_{3}$ from CO$_{2}$ photolysis, which depends on the UV spectral slope of the host star and abundance of H-bearing molecules in the atmosphere \citep{Domagal-Goldman+14, Gao+15, Harman+15, Selsis+02,Tian+14}, 2) massive XUV-driven H-escape and O$_{2}$-build up on planets in the HZ during the pre-main sequence, especially for planets orbiting the latest type stars \citep{Luger+15}, and 3) abiotic O$_{2}$ buildup due to H-escape from N$_{2}$-poor atmospheres that lack tropospheric H$_{2}$O cold traps \citep{Wordsworth+14}\footnote{A more extensive review of oxygen "false positive" mechanisms will be presented in Meadows (2016), in prep.}. In this work we refer to O$_{2}$ and O$_{3}$ produced by these abiotic processes as Òbiosignature impostorsÓ.  Understanding potential false positives for O$_{2}$/O$_{3}$ biosignatures is critical and timely for informing the search for life beyond the solar system, given that the first potentially habitable planets to be characterized with transmission observations will likely orbit M dwarfs, and so will be most susceptible to O$_{2}$ biosignature impostors through CO$_{2}$ photolysis and XUV-driven hydrogen escape (mechanisms (1) and (2) above). 

Possible discriminants for false positive mechanisms have previously been discussed in the literature.  For instance, identifying the byproducts of CO$_{2}$ photolysis using direct imaging was discussed in recent papers that described this false positive mechanism \citep{Domagal-Goldman+14,Harman+15}. N$_{2}$-depleted atmospheres without a cold trap (mechanism (3)) may be identified via the absence of pressure-sensitive N$_{2}$-N$_{2}$ CIA features \citep{Schwieterman+15}. However, previous studies have not addressed identification of CO$_{2}$ photolysis with transmission spectroscopy, or identification of the O$_{2}$-dominated atmospheres that result from massive H-escape. The O$_{2}$ abundance on Earth has self-limited to $\lesssim$ 0.3 bar due to the chemical instability of higher O$_{2}$ abundances with organic matter \citep{Kump08}. Thus, the potentially very high O$_{2}$ pressures from massive H-loss would constitute a separate and discernible regime from biologically produced O$_{2}$ atmospheres.

Here we examine the detectability, in simulated \textit{JWST} transmission spectra, of CO from CO$_{2}$ photolysis as an indicator for abiotically generated  O$_{2}$/O$_{3}$ for a terrestrial HZ planet orbiting an M dwarf. We also examine the detectability of highly pressure-dependent O$_{2}$-O$_{2}$ (O$_{4}$) collisionally induced absorption (CIA) features in transmission and reflectance spectroscopy. These features would be indicators of massive, post-runaway, abiotic O$_{2}$ atmospheres. This examination builds on earlier work by \citep{Misra+14a} that analyzed the capacity of O$_{4}$ bands to determine pressure in O$_{2}$-rich atmospheres, which were presumed to be photosynthetically generated. In \S 2 we describe the models used and the inputs to those models. In \S 3 we present the spectra of the biosignature impostor scenarios and the detectability of our proposed spectral discriminators. We discuss some of the implications of these results in \S 4 and present our conclusions in \S 5. 


\section{Methods and Models}\label{method_model}
	\subsection{Atmosphere Profiles}\label{atm_prof}

	\subsubsection{High-CO photochemical false positive}\label{high_co_atm}
\begin{figure}
   \includegraphics[width=\linewidth]{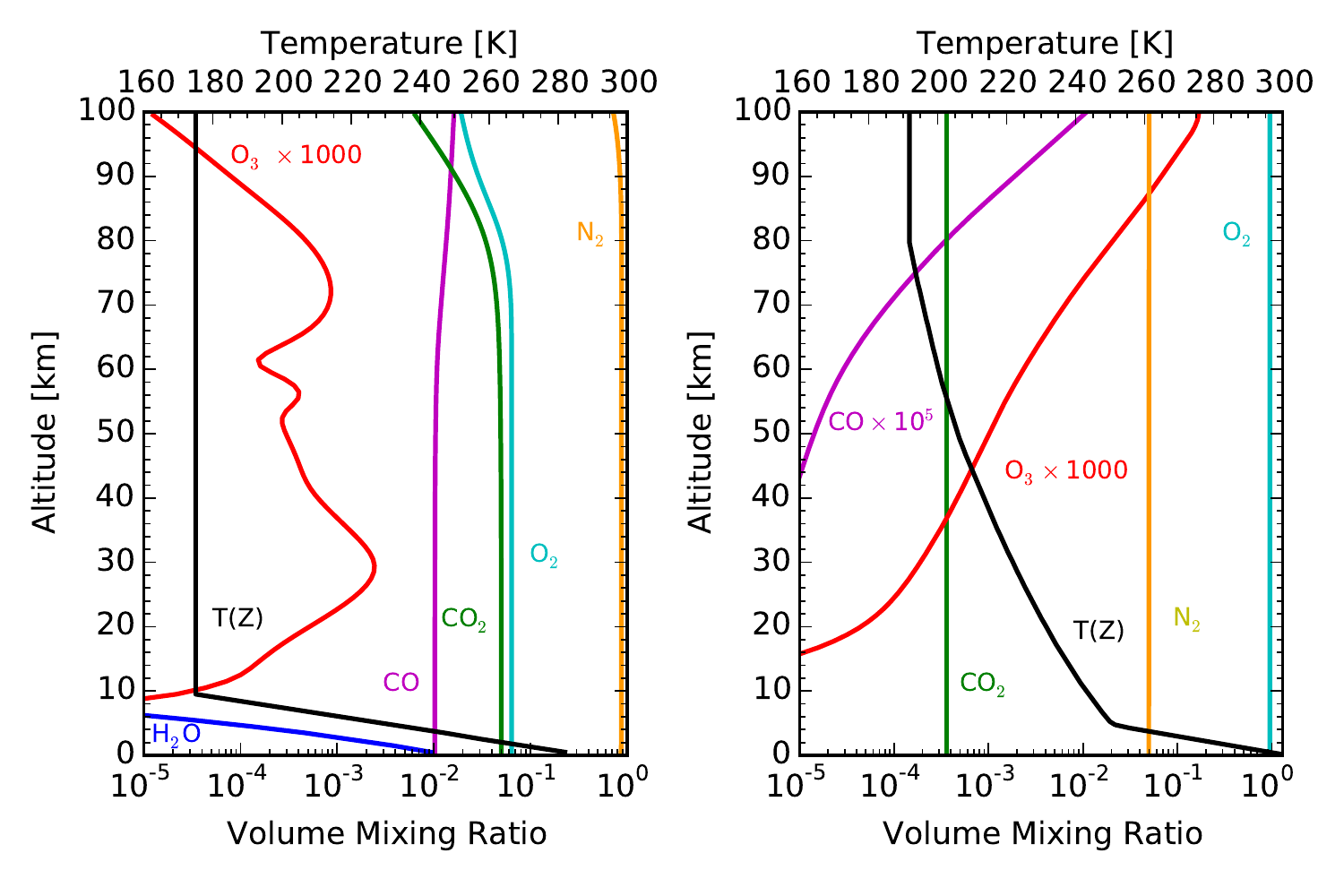}
   \caption{a) temperature and chemical profiles of the GJ 876 Òhigh-O$_{2}$Ó (f$_{\mathrm{O_{2}}}$=6\%) atmosphere from \citet{Harman+15}; b) temperature and chemical profiles for the O$_{2}$-dominated (f$_{\mathrm{O_{2}}}$=95\%), P$_{0}$=100 bar atmosphere calculated for this work.}
   \label{fig1}
\end{figure}

The first atmosphere scenario we consider is a habitable, but lifeless, planet with an N$_{2}$-CO$_{2}$-H$_{2}$O atmosphere, orbiting in the habitable zone of a late (M or K) type dwarf and susceptible to the abiotic accumulation of O$_{2}$ and CO through CO$_{2}$ photolysis:
 \begin{equation}
 \mathrm{CO_{2} + h\nu  (\lambda < 175 nm) \rightarrow CO + O}
 \label{eq:co2_photo}
  \end{equation}
Photochemically liberated O will lead to the formation of both O$_{2}$ and O$_{3}$ through the Chapman scheme, reviewed in \citet{Domagal-Goldman+14}. For this scenario we use the  Òhigh-O$_{2}$Ó case for GJ 876 from \citet{Harman+15} as our atmospheric chemical and temperature profiles (Figure \ref{fig1}a), as that case led to potentially detectable levels of O$_{2}$. GJ 876 is a planet-hosting M4V star with R$_{*}$=0.38 R$_{\odot}$ located 4.66 pc from the solar system \citep{vonBraun+14}. The modeled planet orbits at a semi-major axis of 0.15 AU. Its N$_{2}$-dominated atmosphere contains 5\% CO$_{2}$, 6\% O$_{2}$ and 2\% CO. The temperature profile was calculated in \citet{Harman+15} by assuming a surface temperature commensurate with the stellar insolation and a moist adiabatic lapse rate to an isothermal stratosphere of 175 K. The isothermal stratosphere produces a conservative estimate of the scale height (and therefore transit depths) by neglecting shortwave heating of O$_{3}$. Isothermal stratospheres are commonly chosen in studies of abiotic O$_{2}$ generation (e.g., \citealt{Domagal-Goldman+14, Tian+14}) and we adopt this assumption here to be consistent with previous literature.

\subsubsection{O$_{2}$-dominated post-runaway atmospheres}\label{high_o2_atm}

\citet{Luger+15} calculate that up to thousands of bars of abiotic O$_{2}$ can be generated by XUV-driven H-escape during the pre-main sequence phase of the planet's host star. This depends on a number of factors, including the starting H$_{2}$O inventory (losing the H from one Earth ocean leaves behind up to 240 bars of O$_{2}$). Some of this O$_{2}$ would oxidize the surface; however we assume that there is a point where the oxidization of the planetary surface is complete and O$_{2}$ remains stable in the atmosphere. Even if geological processes slowly draw down this remaining O$_{2}$, much could remain for extended time periods, limited by the rate of mantle overturn. Since many parameters determine how much (if any) abiotic O$_{2}$ would remain behind on these planets when they are observed, we therefore prescribe a range of surface pressures with a 95\% O$_{2}$ mixing ratio: P$_{0}$= 1, 10, and 100 bar. The CO$_{2}$ content is similar to Earth's, f$_{\mathrm{CO_{2}}}$=3.6e-4, and the rest of the atmosphere consists of N$_{2}$ and trace photochemically produced species such as CO and O$_{3}$. To self-consistently calculate temperature and ozone profiles, we used a version of the coupled photochemical-climate model \textit{Atmos}\citep{Arney+16}, based on photochemical and climate models originating with the Kasting group \citep{Kasting+80, Segura+05, Haqq-Misra+08, Domagal-Goldman+14} and upgraded to handle high-pressure O$_{2}$-dominated atmospheres. We assume a completely desiccated planet (no H$_{2}$O) orbiting at the inner edge of the HZ (a = 0.12 AU) of a star with identical properties to GJ 876. See Figure \ref{fig1}b for the atmospheric chemical profiles of the P$_{0}$=100 bar scenario. 

\subsection{Radiative Transfer Models and Inputs}\label{radiative_transfer}
The core radiative transfer code used for this work is the Spectral Mapping Atmospheric Radiative Transfer (SMART) model developed by D. Crisp \citep{Crisp97, Meadows+96}, which is a line-by-line, multi-stream, multi-scattering model. SMART is well-validated from data-model comparisons of Earth, Venus, and Mars (e.g., \citealt{Robinson+11, Robinson+14, Arney+14}). SMART can be used to calculate synthetic direct-imaging spectra and transmission spectra using a transmission capable version of SMART that includes refraction \citep{Misra+14a,Misra+14b}. All modeled spectra include O$_{4}$ absorption (O$_{2}$-O$_{2}$ CIA), although it only produces a strong spectral impact in the cases with O$_{2}$-dominated atmospheres. Weak O$_{4}$ bands are present in Earth's disk-averaged spectrum \citep{Tinetti+06,Turnbull+06}. O$_{4}$ absorption has been found to be only very weakly temperature-dependent, suggesting that O$_{2}$-O$_{2}$ CIA dominates over true van der Waals molecule absorption \citep{Thalman+13}.  The CIA absorption from O$_{4}$ varies quadratically with density, and the absorption coefficients can be given as:
\begin{equation}
\mathrm{\alpha(\lambda,T,d_{O_{2}}) = B_{O_{2}-O_{2}}(\lambda,T)d^{2}_{O_{2}} \approx B_{O_{2}-O_{2}}(\lambda)d_{O_{2}}^{2}}
\label{eq:o4_abs}
\end{equation}
where $\alpha$ is the absorption coefficient, d$_{\mathrm{O_{2}}}$ is the density of O$_{2}$ molecules, and B$_{\mathrm{O_{2}-O_{2}}}$ is a density-normalized CIA coefficient. We use the density-normalized O$_{4}$ coefficients from C. Hermans\footnote{http://spectrolab.aeronomie.be/o2.htm}  for the 0.333-0.666 $\mu$m spectral range \citep{Hermans+99}, and the values from \citet{Greenblatt+90} and \citet{Mate+99} for the 1.06 and 1.27 $\mu$m features. 

\subsection{JWST NIRISS \& NIRSpec noise model}\label{noise_model}
We calculated the synthetic \textit{JWST} observations as described by \citet{Deming+09} with some updates.  The total throughput of the NIRISS \citep{Doyon+14} and NIRSpec  \citep{Ferruit+14} instruments, including the telescope, were taken from Albert (2015, private communication) and from http://www.cosmos.esa.int/web/jwst/nirspec-pce, respectively.  Thermal and zodiacal background were included as described by \citet{Deming+09}, but the background only becomes significant longward of approximately 4 $\mu$m.  There is very little overhead for NIRISS transit spectroscopy (Albert 2015, private communication), so we adopted an observing efficiency of unity for NIRISS.  For NIRSpec, we calculated the observing efficiency assuming a 2048 $\times$ 32 subarray \citep{Tumlinson+10}, and adopted a number of groups in an integration (ngrp) of 4 from \citet{Karakla+10}.  For both instruments, the observations are close to saturation, but we assume that saturation can be avoided by slightly dithering or scanning the telescope perpendicular to dispersion.  We adopt photon-limited noise, dropping the hypothetical instrument systematic noise used by \citet{Deming+09}.  Recent HST experience has demonstrated close to photon-limited performance for bright stars \citep{Kreidberg+14}. We represent the star using a Phoenix model atmosphere \citep{Allard+95}.


\section{Simulated Spectra of Atmospheres with Abiotic O$_{2}$}\label{simulated_spectra}

We present photochemically self-consistent test cases for two different false positive mechanisms: 1) an Earth-size planet with a prebiotic atmosphere (Figure \ref{fig1}a) orbiting in the HZ of GJ 876, and 2) an Earth-size planet with an O$_{2}$-dominated (95\%) 100-bar atmosphere orbiting at the inner edge of the HZ of GJ 876 (Figure \ref{fig1}b). These atmospheres are free of clouds and aerosols. We examine scenario (2) in transmitted and reflected light. In the reflected light case we treat GJ 876 as a stand-in for the handful of nearby late-type stars whose habitable zones may be examined with future direct-imaging telescopes such as HDST/LUVOIR \citep{Stark+14, Dalcanton+15}. 

\begin{figure*}
   \includegraphics[width=\linewidth]{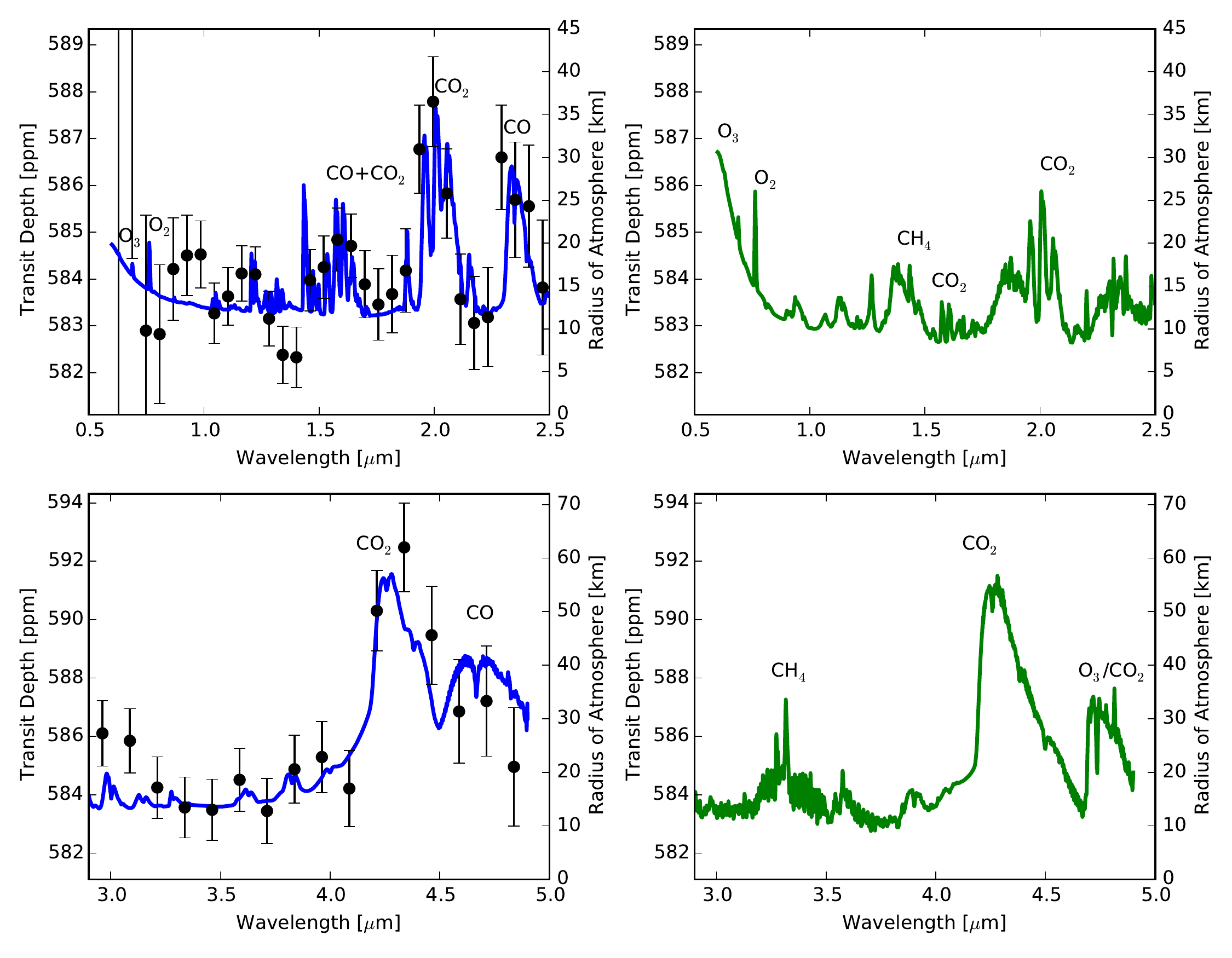}
   \caption{Left: Spectrum (blue) of photochemical high-O$_{2}$/CO GJ 876 atmosphere from \citet{Harman+15} in the \textit{JWST}-NIRISS band (top) and in the \textit{JWST}-NIRspec band (bottom)). Data points and 1$\sigma$ error bars were generated with the \textit{JWST} instrument simulator \citep{Deming+09} assuming 65-hour integrations (10 transits of GJ 876) and photon-limited noise. Right: Comparable model spectrum (green) of Earth orbiting GJ 876 using atmosphere profiles taken from Figure 1 of \citet{Schwieterman+15}.}
   \label{fig2}
\end{figure*}

Figure \ref{fig2} shows the calculated spectral transmission depths and simulated observations of case (1) in the \textit{JWST} NIRISS (0.6-2.5 $\mu$m) and NIRSpec (2.9-5.0 $\mu$m) bands (left panel). Our calculations show that the 1.65, 2.0, and 4.3 $\mu$m CO$_{2}$ bands can be detected with a SNR $>$ 3 (3.1, 4.3 and 8.0, respectively) when binned across the absorption bands and compared to the continuum level. The 2.35 $\mu$m CO band has a SNR of 3.7 while the 4.6 $\mu$m CO band has an SNR of 2.6. Other absorption bands have SNRs $\lesssim$ 1. The simultaneous detection of both CO$_{2}$ and CO could indicate CO$_{2}$ photolysis in the planet's atmosphere. Note that for the integration time used here, the relatively narrow O$_{2}$-A band (0.76 $\mu$m) and weaker O$_{2}$ features would not be detectable. On the right side panel of Figure \ref{fig2} we show a spectrum of modern Earth orbiting GJ 876 using atmosphere profiles from Figure 1 of \citet{Schwieterman+15}, but with otherwise the same parameters as the left panel. This spectrum is not self-consistent with the star, but provided for comparison.

\begin{figure}
   \includegraphics[width=\linewidth]{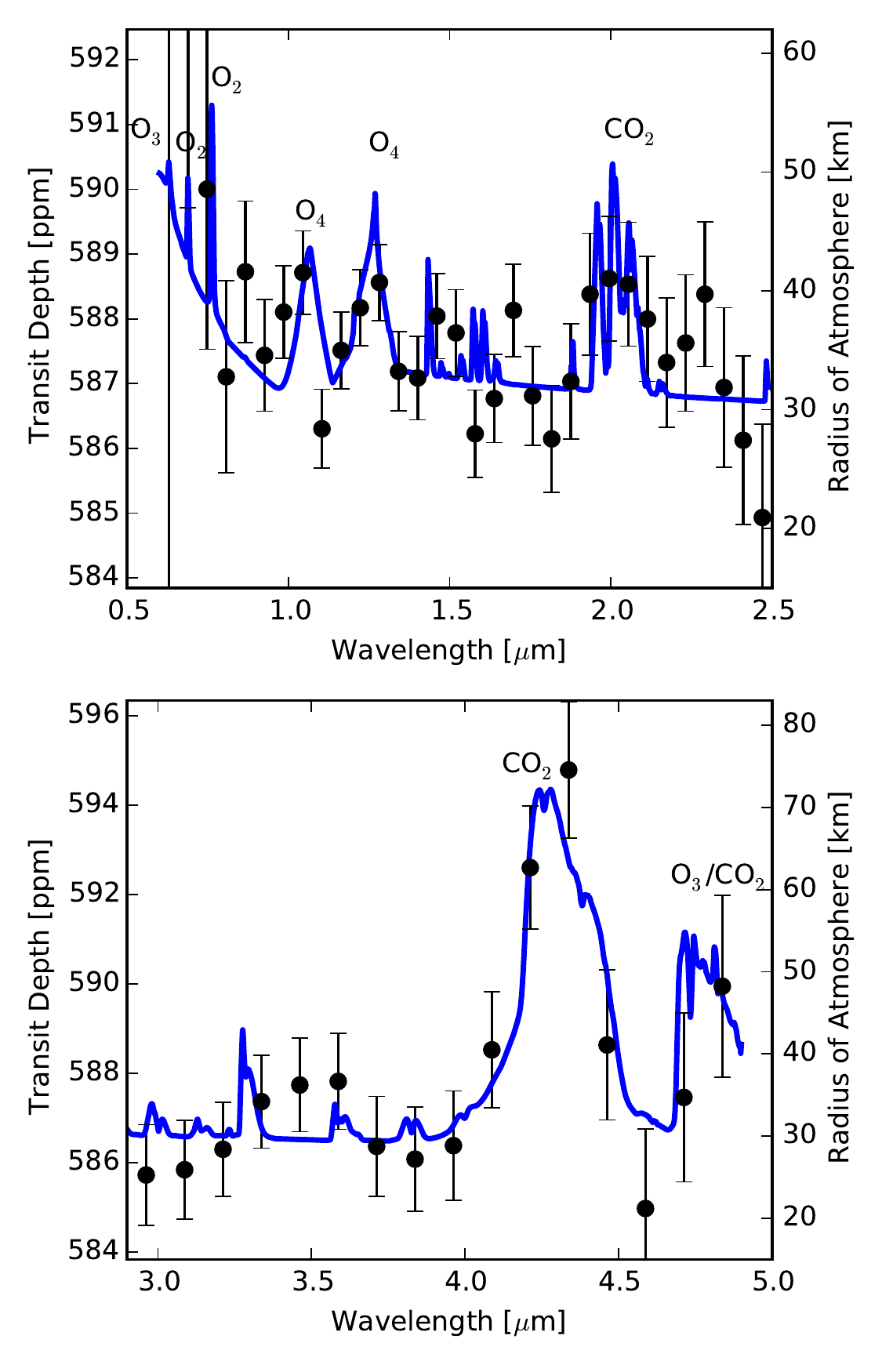}
   \caption{Spectra (blue) of 100 bar post-runaway O$_{2}$ atmosphere in \textit{JWST}-NIRISS band (top) and in the \textit{JWST}-NIRSpec band (bottom). Data points and 1$\sigma$ error bars were generated with the \textit{JWST} instrument simulator \citep{Deming+09} assuming 65-hour integrations (10 transits of GJ 876) and photon-limited noise. See right panels of Figure \ref{fig2} for a comparable Earth spectrum.}
   \label{fig3}
\end{figure}

Figure \ref{fig3} shows the calculated transmission spectrum depths of the high-O$_{2}$ atmosphere from case (2). The SNRs of the 1.06 and 1.27 $\mu$m O$_{4}$ bands are 2.8 and 3.1, respectively. The detection of strong O$_{4}$ bands would indicate a very massive O$_{2}$ atmosphere free of aerosols at high altitudes. 

\begin{figure}
   \includegraphics[width=\linewidth]{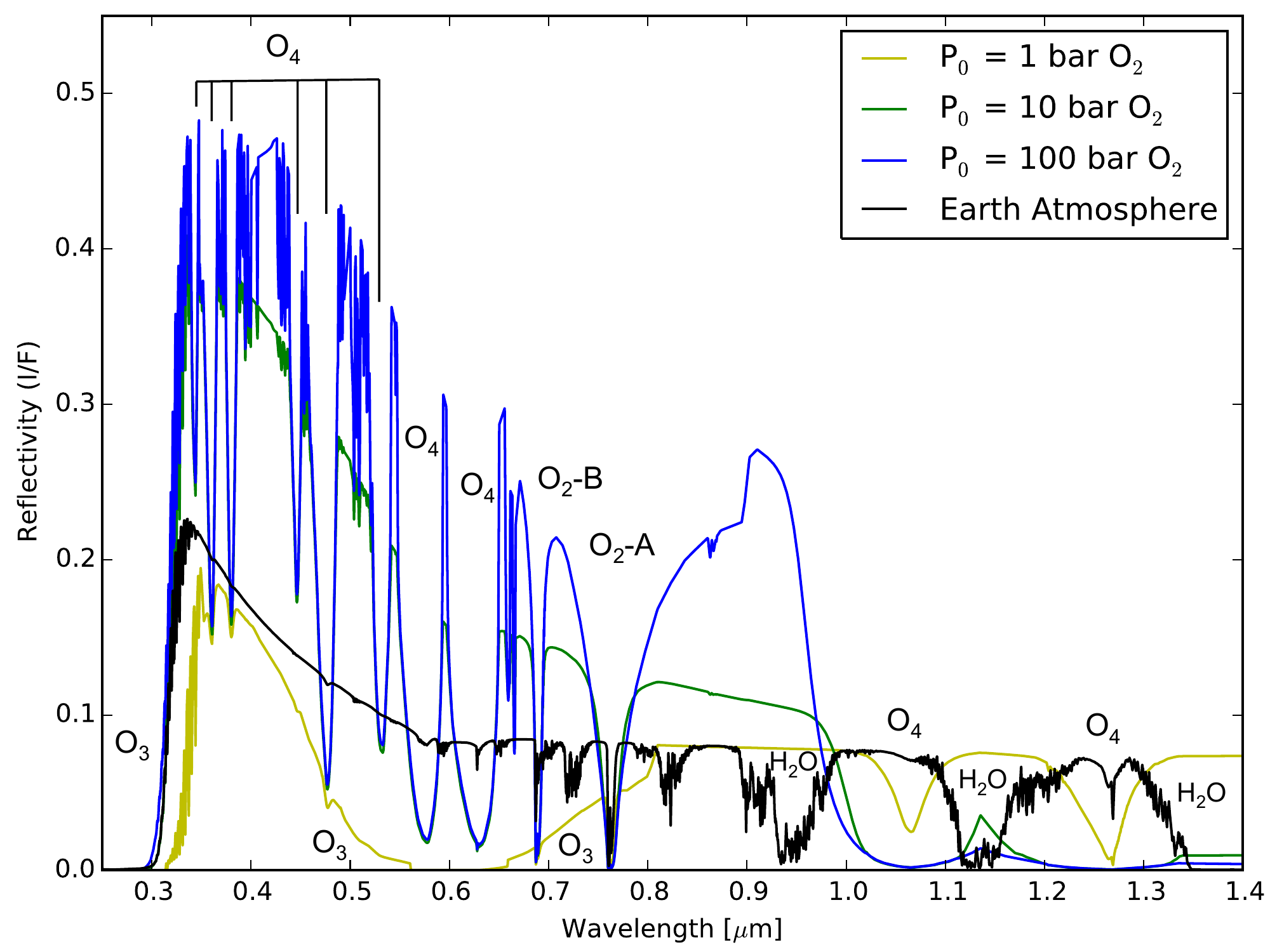}
   \caption{Synthetic reflectance spectra of 1, 10, and 100 bar high-O$_{2}$ atmospheres (yellow, green, and blue, respectively) with O$_{2}$ and O$_{4}$ bands identified. A comparable Earth spectrum is shown in black.}
   \label{fig4}
\end{figure}

In Figure \ref{fig4}, we show simulated reflectance (direct-imaging) spectra of O$_{2}$-dominated atmospheres (P$_{0}$=1, 10, and 100 bar) seen at quadrature (solar zenith angle of 60$\degr$) using SMART. We use the chemical profiles described in \S 2.1.2. For comparison we include the spectrum of an Earth atmosphere \citep{Schwieterman+15}.   A gray surface albedo of A$_{B}$ = 0.15 is assumed. The O$_{4}$ bands at 0.345, 0.36, 0.38, 0.445, 0.475, 0.53, 0.57, 0.63, 1.06, and 1.27 $\mu$m are labeled, with the NIR bands saturating at the highest O$_{2}$ abundance. This simple test case demonstrates that if the high-O$_{2}$ atmospheres proposed by \citet{Luger+15} exist, the O$_{4}$ absorption band strength in those planetary spectra would rival or exceed that of the monomer O$_{2}$ bands. These spectra are qualitatively different than modern-Earth's spectrum, even in the 0.3-1.0 $\mu$m range, with a different shape, broader O$_{2}$ features, and additional features from O$_{4}$. These are all signs of a much higher O$_{2}$ abundance than the Earth's atmosphere - self-regulated by negative feedbacks - has ever achieved. \hfill \break
\hfill \break


\section{O$_{4}$ and CO as Spectral Indicators of Abiotic Oxygen}\label{discussion}
The first potentially habitable terrestrial planet atmospheres to be characterized via transmission spectroscopy by \textit{JWST} or large ground-based observatories will likely be orbiting near the inner edge of the HZ of late type stars \citep{Deming+09, Sullivan+15}. This is due to the detection biases of planet transit searches and the shorter orbital periods of HZ planets around late type stars, allowing greater potential to integrate over multiple transits for characterization. Unfortunately, according to current modeling studies \citep{Harman+15, Luger+15} these planets are the worlds in which Òbiosignature impostorsÓ- abiotic O$_{2}$ and O$_{3}$ - are most likely. Furthermore, this potential for false positives for life is even greater for the extended inner habitable zone for ÒdryÓ planets \citep{Abe+11}, which may have an even greater probability of being characterized first. The development of observing strategies to mitigate the potential for a Òfalse positiveÓ biosignature detection is thus relevant now. We argue that the detection of significant CO and CO$_{2}$ could indicate robust CO$_{2}$ photolysis, which can produce abiotic O$_{2}$/O$_{3}$, and also that strong O$_{4}$ bands would be indicative of an oxygen atmosphere too massive to be biological. Thus CO$_{2}$/CO and O$_{4}$ are potentially powerful spectral discriminators against abiotic O$_{2}$/O$_{3}$ in future potentially habitable exoplanet observations. 

Furthermore, in the cases presented here, the spectral discriminators against abiotic  O$_{2}$/O$_{3}$  are more detectable with a hypothetical \textit{JWST} observation than the O$_{2}$ or O$_{3}$ signatures themselves. In our example spectra, neither O$_{2}$ nor O$_{3}$ would be directly detectable with just 10 transits, but the abiotic discriminators CO/CO$_{2}$ and O$_{4}$ could be. This provides an opportunity to maximize the utility of observing time if the ultimate goal is to characterize planets where true biosignatures are obtainable. If spectral indicators for biosignature impostors are detected with reasonable confidence, the community may wish to reallocate the remaining time to other promising targets, rather than integrate further. Additionally, it may also be the case that O$_{2}$ or O$_{3}$ may be identified in the same targets via other observing strategies at a concurrent or future date. For example, it has been proposed that O$_{2}$ may be found via ground-based Extremely Large Telescope (ELT) observations of the 0.76 $\mu$m O$_{2}$-A \citep{Snellen+13}. Observations by space-based telescopes such as \textit{JWST} or a future LUVOIR telescope operating in transit mode on a favorable target could detect or rule out these indicators before significant observing time is expended by ELTs on promising targets. If an ELT observation of a potentially habitable planet is conducted, these potential spectral discriminators could be characterized in addition to the target band (e.g., the O$_{2}$-A band).  

It is important to note that the strength of an absorption band in transmission, unlike in reflected light, is not dependent on the mixing ratio of the gas at the surface or the absolute column abundance of the gas. Rather, the strength is mostly dependent on the altitude at which the gas produces an optical depth near unity, and thus the distribution of the gas in the atmosphere is extremely important. Direct imaging of CO$_{2}$ and CO spectral signatures on an Earth twin is potentially problematic because they are weak and narrow shortward of 2.0 $\mu$m and the planetary flux at the strongest bands (e.g., the 4.3 $\mu$m CO$_{2}$ band) is low \citep{Robinson+14, Schwieterman+15}. Our work shows these bands are much stronger in transmission.  Thus transmission spectroscopy is complementary to potential future direct-imaging characterization missions, and in particular for characterizing potential biosignature impostors.  Transmission spectroscopy is also more sensitive to Earth-like CH$_4$ abundances than direct-imaging. Detection of CH$_4$ with O$_2$/O$_3$ would help confirm a true biosignature, as modeling predicts extremely low CH$_4$ abundances in cases of abiotic O$_2$/O$_3$  generation \citep{Domagal-Goldman+14}. 

Next-generation space-based direct-imaging telescope concepts such as ATLAST/LUVOIR, HDST, and HabEx would be able to directly image planets in the HZ \citep{Rauscher+15, Dalcanton+15, Swain+15}, and thus potentially characterize biosignature gases in exoplanet atmospheres. Detection of strong O$_{4}$ bands in UV/VIS/NIR reflected light would indicate a large O$_{2}$ atmosphere originating from massive H-escape. While planets in the ÒconservativeÓ HZ of G and F type stars would be high priority for these missions and are less susceptible to an extended history of runaway H-loss and O$_{2}$-buildup, planets orbiting between the ÒoptimisticÓ (recent Venus) and ÒconservativeÓ (runaway greenhouse) inner edge of the HZ, potentially brighter targets for imaging, would also be susceptible to this process \citep{Luger+15}.  Additionally, many nearby K and M type stars would also likely be characterized \citep{Stark+14, Dalcanton+15}, where larger regions of the HZ would be susceptible to generating biosignature impostors. 


\section{Conclusions}\label{conclusions}
Recently proposed mechanisms for developing abiotic O$_{2}$/O$_{3}$ in terrestrial exoplanet atmospheres would produce spectral discriminators that are potentially identifiable with future telescope observations, including \textit{JWST}. These discriminants are more detectable than O$_{2}$ or O$_{3}$ in transmission observations.  CO seen at 2.35 $\mu$m or 4.6 $\mu$m with CO$_{2}$ at 2 or 4.3 $\mu$m would indicate robust CO$_{2}$ photolysis and suggest a high likelihood of abiotic O$_{2}$/O$_{3}$ generation. We find that CO in a realistic exoplanet atmosphere orbiting a late type star could be seen with a SNR $>$ 3 at 2.35 $\mu$m in \textit{JWST}-NIRISS with as few as 10 transits. O$_{4}$ bands seen in transmission or direct-imaging can be diagnostic of high-O$_{2}$ post-runaway atmospheres that have experienced a history of H-escape. The 1.06 and 1.27 $\mu$m O$_{4}$ bands in a massive, O$_{2}$-dominated atmosphere without high-altitude aerosols would be potentially detectable with a SNR $\gtrsim$ 3 with as few as 10 transits with \textit{JWST}-NIRISS, assuming photon-limited noise. 


\acknowledgements
This work was supported by the NASA Astrobiology Institute's Virtual Planetary Laboratory Lead Team, funded through the NASA Astrobiology Institute under solicitation NNH12ZDA002C and Cooperative Agreement Number NNA13AA93A. This research used the advanced computational, storage, and networking infrastructure provided by the Hyak supercomputer system at the University of Washington. This work made use of the NASA Astrophysics Data System. We would like to thank the anonymous reviewer for helpful comments, which improved the manuscript.



                                                                                   
\end{document}